\documentstyle[prl,twocolumn,aps]{revtex}
\begin{document}
\draft

\title{One-dimensional Josephson arrays as
superlattices for single Cooper pairs
         }
\author{
A. A. Odintsov
}
\address{
Department of Applied Physics, Delft University of Technology,
2628 CJ Delft, The Netherlands \\
and  Nuclear Physics Institute, Moscow
State University, Moscow 119899 GSP, Russia \\
}
\address{\rm  (Received 10 October 1995)}
\address{\mbox{ }}

\address{\parbox{14cm}{\rm \mbox{ }\mbox{ }\mbox{ }
We investigate uniform one-dimensional arrays of small Josephson
junctions ($E_J \ll E_C$, $E_C = (2e)^2/2C$)
with a realistic  Coulomb interaction
$U(x) = E_C \lambda \exp( - |x|/\lambda)$
(here  $\lambda \gg 1$ is the screening length
in units of the lattice constant of the array).
At low energies this system can be described in terms of interacting
Bose particles (extra single Cooper pairs) on the lattice.
With increasing concentration $\nu$ of extra Cooper pairs,
a crossover from the Bose gas phase
to the Wigner crystal phase and then to the superlattice
regime occurs.
The phase diagram in the superlattice regime
consists of commensurable insulating phases with
$\nu = 1/l$ ($l$ is integer)
separated by superconducting regions
where the current is carried by excitations with
{\em fractional} electric charge $q = \pm 2e/l$.
The Josephson current through a ring-shaped array
pierced by magnetic flux is calculated for all of the phases.
}}
\address{\mbox{ }\mbox{ }}
\address{\parbox{14cm}{\rm PACS numbers:
05.30.Jp, 73.40.Gk, 74.50.+r, 73.20.Dx, 72.15.Nj}}

\maketitle

\tighten
\narrowtext

The investigation of arrays of small Josephson junctions
attracts growing interest of theoreticians and experimentalists
(see \cite{Mooij} for a review).
In such arrays the Coulomb energy $E_C = (2e)^2/2C$
can be of the order of the Josephson energy $E_J$
(here $C$ is the capacitance of the junctions).
Since the Josephson phase $\phi$ and
the electric charge $Q$ on the islands
are canonically conjugated operators, $[\phi, Q]  = 2ei$,
a number of nontrivial quantum phenomena arise due to
a competition between the phase (or vortex) and charge
degrees of freedom.
In particular, the point of the
superconductor-insulator transition depends on
the magnetic frustration~\cite{msi}
and on the electro-chemical potential $\mu$ of the
array~\cite{esi,Freer,esilong,Batr}.

Existing theories of electric field-induced
su\-per\-con\-duc\-tor-\-in\-sulator transition
\cite{esi,Freer,esilong,Batr} treat
predominantly the cases of on-site or nearest-neighbor
Coulomb interaction, although for typical experimental parameters
the range of the interaction is large~\cite{Mooij},
$\lambda = 3 \div 20$
(in units of the lattice constant of the array).
The phase diagram of the superconductor-insulator transition
becomes rather complicated for $\lambda \gg 1$,
{\it and} large concentrations $\nu \sim 1/\lambda^d$ of
field-induced ("extra") Cooper pairs~\cite{esilong}
(here $d$ is a dimension of the array).
This is related to the fact that
the extra Cooper pairs (ECP)
can form a variety of configurations commensurable with
the lattice of junctions.
On the other hand, it is known \cite{OdNaz} that
at very small concentrations~\cite{remark1} $\nu \ll 1/\lambda^d$
the ECP form a Bose gas with hard-core interaction
and commensurability with the lattice plays no role.
In this work we study the transition
between these two very different regimes.

We consider uniform one-dimensional (1-D) Josephson arrays
with large Coulomb energy $E_C \gg E_J$
and long range Coulomb interactions, $\lambda \gg 1$.
We focus on  the regime with {\it low concentrations}
$\nu \ll 1/\lambda$ of ECP.
We will show that the phase diagram of
the superconductor-insulator transition
has simple structure in this regime.
In the insulating phase the ECP
form a regular superlattice with the period $l$  ($\nu = 1/l$).
The system starts showing superconducting
properties when the first mobile excitation appears.
The latter corresponds to a change of
the distance between two neighboring ECP in a superlattice by
$\Delta l = \pm 1$.
This excitation can be treated as a quasiparticle
with fractional charge $q = \mp 2e/l$.
The superconducting phase can be viewed a gas of such
quasiparticles on the lattice.

The Hamiltonian  $H = H_C + H_J$ of 1-D Josephson array
consists of a Coulomb term $H_C$ and a Josephson term $H_J$.
The Coulomb energy is given by
$H_C = \frac{1}{2}  \sum_{i,j=1}^{L} n_i U(i-j) n_j - \mu N$,
where $n_i$ is a (positive or negative) number of ECP
on the electrode $i$,
$N = \sum_{i=1}^{L} n_i$, and $U(x) = E_C \lambda \exp (-x/\lambda)$
is the Coulomb interaction between ECP. The interaction is screened
at the length $\lambda = (C/C_g)^{1/2}$
due to a finite self-capacitance $C_g$ of the islands.
The gate voltage $V$
plays the role of the chemical potential, $\mu = 2eV$.

We consider first the limit of zero Josephson coupling.
For $| \mu | < \mu_{tr} \equiv \lambda E_C/2$
there is no ECP in the array ($n_i = 0$ for all $i$).
Just above the threshold,
$0 < \mu - \mu_{tr} \ll \mu_{tr}$,
the ground state is still characterized by the absence of ECP
on most of the islands ($n_i = 0$).
The rest of the islands are occupied by one ECP ($n_i = 1$).
The configurations with $n_i \neq 0, 1$ (for some $i$)
are separated from the ground state by a Coulomb gap
$\Delta_C = E_C/\lambda$,
which corresponds to the difference between
the Coulomb energies of the configurations
$\{ n_i \} = (0,...,1,-1,1,0,...)$ and
$(0,...,1,0,...)$.
In what follows we restrict the space of states
to low energy configurations with  $n_i = 0$ or $1$.
These configurations can be fully characterized by
the coordinates $x_j$ of ECP on the lattice
($x_j$ are integer numbers).

The Josephson term of the Hamiltonian has a standard form,
$H_J = - E_J \sum_{i=1}^{L} \cos(\phi_{i+1} - \phi_i -a)$.
Here  $\phi_j$ is the operator of the Josephson phase
of the island $j$
obeying the commutation relation
$[ n_j, \phi_{j'} ] = - \delta_{j,j'} i$, and
$a = (2 \pi / L) \Phi/\Phi_0$ is the vector potential
(for circular array pierced by a magnetic flux $\Phi$).
The Josephson term acting on the restricted space of states
describes a hoping of ECP on the lattice with
the amplitude $E_J e^{\pm i a}/2$.
Corrections to the tunneling amplitude due to the
states with $n \neq 0, 1$ are small for $E_J \ll \Delta_C$.
Therefore, the original Hamiltonian $H$
in the low-energy space is equivalent to the
Hamiltonian of Bose particles (ECP) on the lattice~\cite{remfermions},
\begin{equation}
H = - E_J \sum_{j=1}^{N} \cos(p_j - a)
   +    \sum_{j=1}^{N} U(x_{j+1} - x_j) - \bar{\mu} N,
\label{ham}
\end{equation}
where $p_j$ are quasimomenta of ECP and $\bar{\mu} = \mu -\mu_{tr}$.
Since the interaction potential $U(x)$
is screened strongly on the interparticle distances
$\nu^{-1} \gg \lambda$
we take into account the interaction of the neighboring ECP only.

\begin{figure}
\vbox to 3.5cm {\vss\hbox to 8cm
 {\hss\
   {\includegraphics{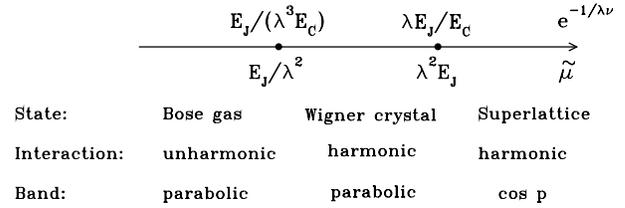} }
  \hss}
}
\caption{Schematic phase diagram of 1-D Josephson array.
\label{phdqual}}
\end{figure}

Let's consider now the ground state of the system as a function
of the chemical potential $\bar{\mu}$ at fixed $E_J$.
The first ECP enters the array at
$\bar{\mu} = - E_J$
We denote the deviation from this threshold by
$\tilde{\mu} = \bar{\mu} + E_J$.
At low concentrations $\nu$ (see Eq.(\ref{nubose}))
ECP form a Bose gas with hard core interaction~\cite{OdNaz}
(Fig.~\ref{phdqual}).
The chemical potential $\tilde{\mu}^{(BG)} = (\pi^2 / 2) E_J \nu^2 $
is determined by the kinetic energy of the particles in parabolic band
\cite{Lieb}.
Taking into account a finite range of the interaction we obtain
by a variational calculation that
the correction to $\tilde{\mu}^{(BG)}$
due to a finite size $d = \lambda \ln(\lambda^3 E_C / E_J)$
of the core is small for
\begin{equation}
\nu \lambda \ln( \lambda^3 E_C/E_J) \ll 1.
\label{nubose}
\end{equation}

With increasing concentration ECP get localized
in coordinate space.
We assume {\it apriori} that ECP form 1-D Wigner crystal.
Expanding kinetic and potential energy (\ref{ham})
up to quadratic terms we obtain the chemical potential
$
\tilde{\mu}^{(WC)} =
(2/\pi) (E_C E_J / \lambda)^{1/2} \exp (- 1/ 2 \lambda \nu)
 + \lambda E_C \exp (-1/ \lambda \nu)
$
and the mean-square displacement of neighboring particles
$\langle (x_{i+1} - x_i)^2 \rangle = [2E_J/U_0 (\nu^{-1})]^{1/2} / \pi$,
where
\begin{equation}
U_0(l)  \equiv  (1/2) d^2 U(x) /d x^2 |_{x=l} =
(E_C/2\lambda) \exp(-l/\lambda).
\label{U_0}
\end{equation}
Note that the concentration of ECP $\nu(\tilde{\mu})$
increases much more slowly than in the Bose gas regime.
This expansion is legitimate if
the fluctuations of displacement are small,
$\langle (x_{i+1} - x_i)^2 \rangle \ll \lambda^2$,
and the kinetic energy per ECP is much less than $E_J$.
These conditions determine
respectively
the lower and the upper bound of the range of concentrations
\begin{equation}
\frac{E_J}{\lambda^3 E_C} \ll
\exp \left( - \frac{1}{\nu \lambda} \right) \ll
\frac{\lambda E_J}{E_C}
\label{WC}
\end{equation}
in which the Wigner crystal phase exists (Fig.~\ref{phdqual}).
For both the phases (Bose gas and Wigner crystal)
the Hamiltonian is quadratic in
momenta of the particles.
Therefore, the vector potential $a$ is coupled
to the  momenta
\noindent
of the center of mass only.
For this reason the
Josephson current through a ring-shaped array
is given by a universal expression
\begin{equation}
I_J = \frac{4ev_J}{L} \frac{\Phi}{\Phi_0}, \; v_J = \pi \nu E_J,
\label{Ijharm}
\end{equation}
(for $|\Phi| < \Phi_0/2$).
A similar result holds for the persistent current of interacting
Fermions~\cite{Loss} $v_J$ being the Fermi velocity in that case.

\begin{figure}
\vbox to 11cm {\vss\hbox to 8cm
 {\hss\
   {\includegraphics{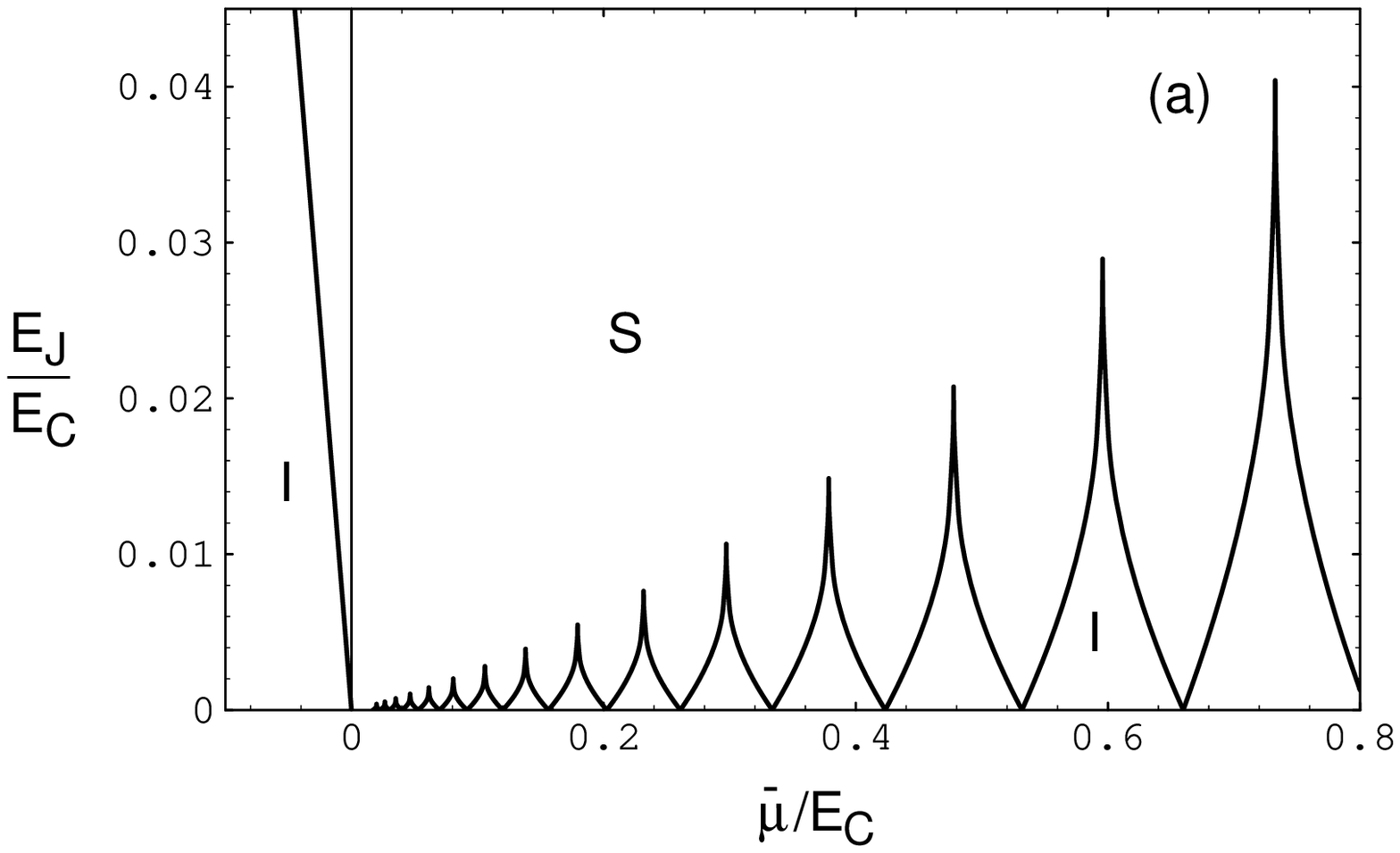}
    \includegraphics{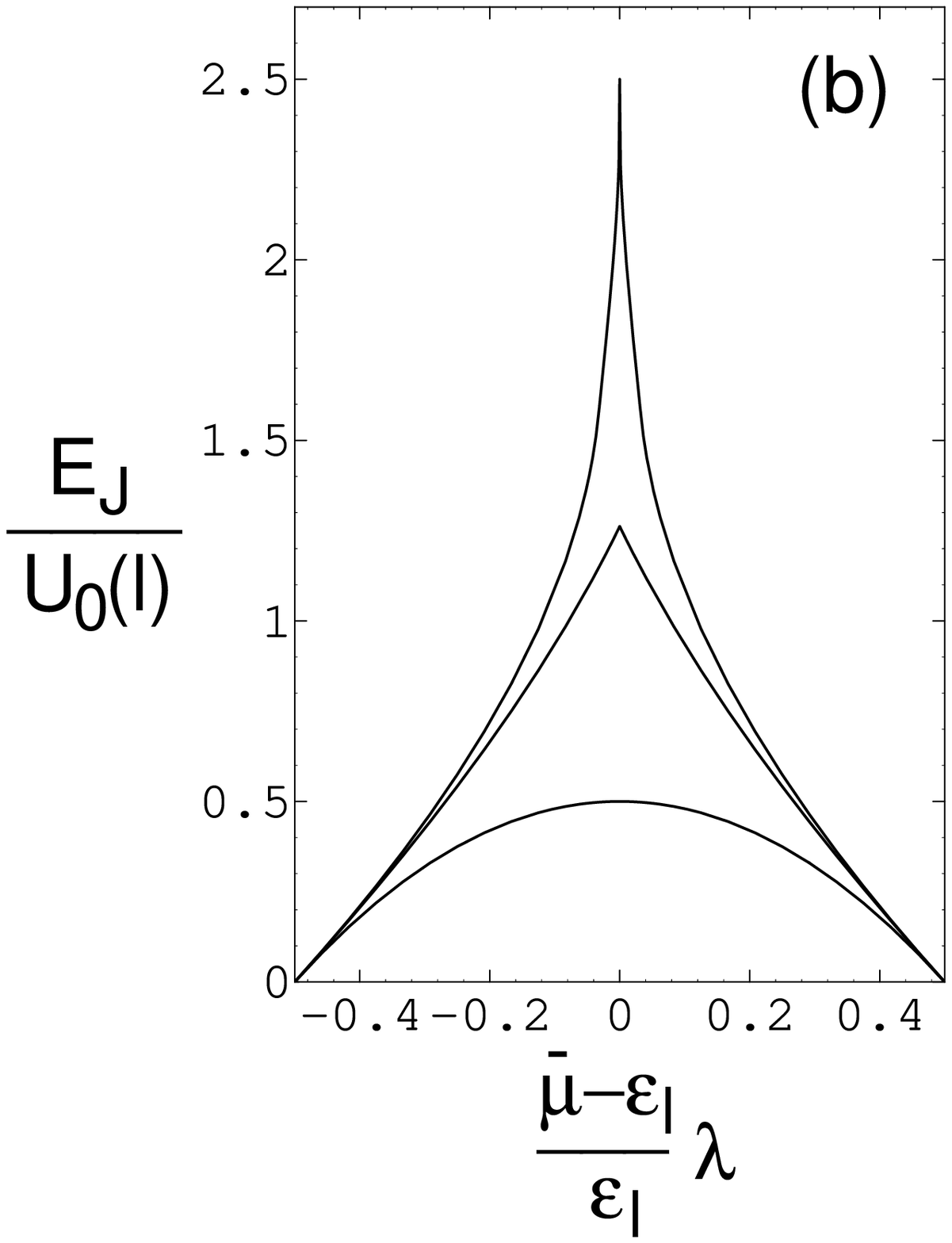}
    \includegraphics{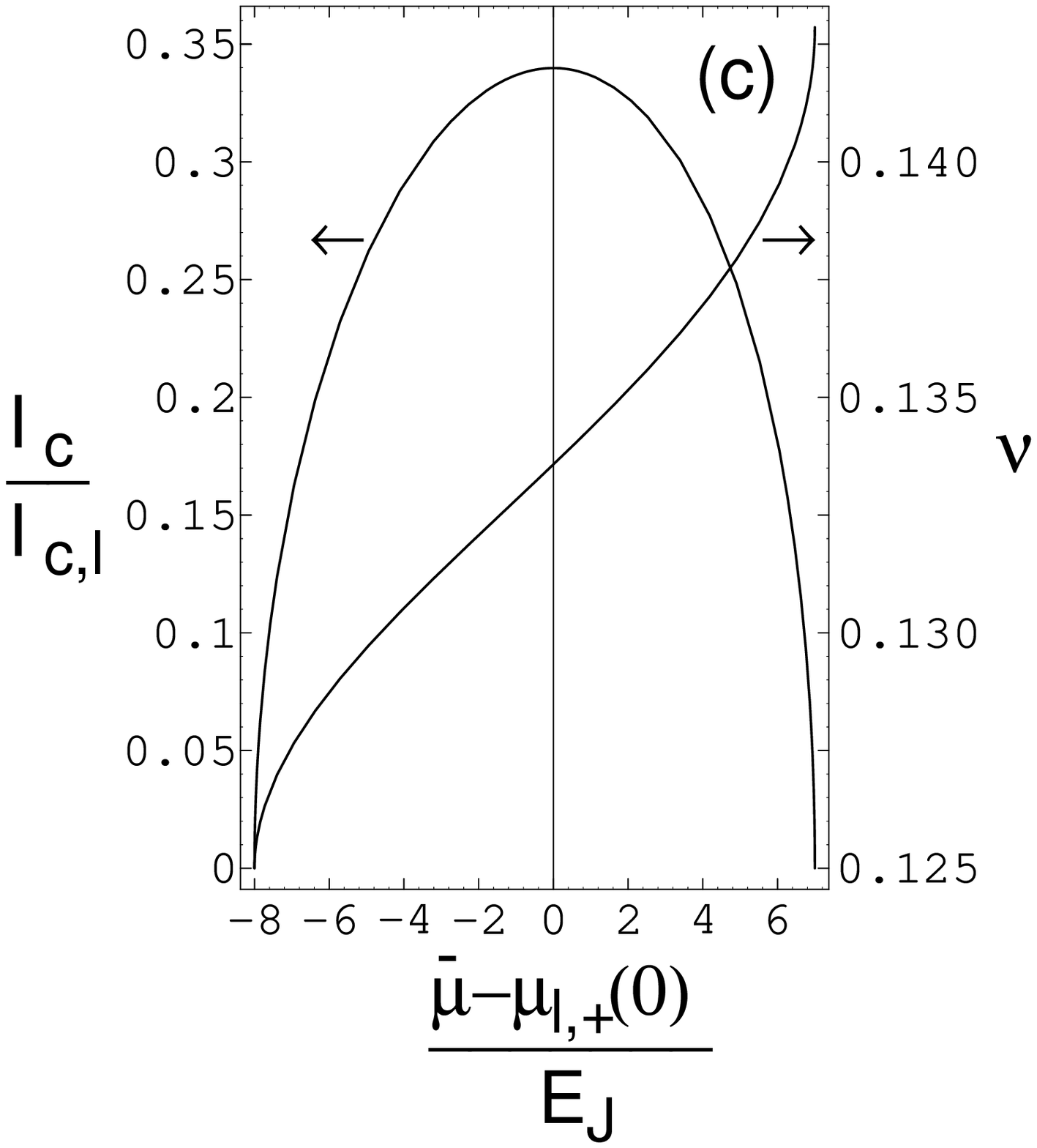}
   }
  \hss}
}
\caption{Phase diagram of the
superconductor-insulator transition
in 1-D Josephson array ($\lambda = 3$).
(a) The overall view.
The insulating spike-like regions from right to left
correspond to $\nu = 1/7, 1/8,...$.
(b) Boundary of the insulating phase with $\nu = 1/l$ ($l \gg 1$).
The curves from bottom to top correspond to
the results of the mean-field approach,
the third order perturbation theory,
and the extrapolation of the perturbation theory to
infinite order connected with the extrapolation of
the Kosterlitz-Thouless scaling.
(c) Concentration $\nu(\bar{\mu})$ of ECP
and critical Josephson current $I_c(\bar{\mu})$
in the superconducting region $1/8 < \nu < 1/7$ ($l = 8$).
We normalize $I_c$ by the critical current (5)
for the concentration $\nu = 1/l$ of  ECP, $I_{c,l} = 2\pi e E_J/lL$. }
\label{phd}
\end{figure}

For larger concentrations of ECP the kinetic energy per particle
becomes comparable to the bandwidth $E_J$.
To investigate this case we start from the limit
$E_J \to 0$.
First we determine the range of the chemical potential
$\mu_{l,-}(0) < \bar{\mu} < \mu_{l,+}(0)$
where the configuration with
equidistantly spaced (at a distance $l$)
ECP is the ground state of the system.
In order to add (subtract) a Cooper pair into (from) this configuration,
one has to convert $l$ bonds of length $l$ between neighboring ECP
into $l$ bonds of length $l-1$ ($l+1$).
The energy required for this conversion is given by
$\mu_{l,\pm}(0) = \pm l [ U (l \mp 1) - U (l) ] \simeq
\epsilon_l (1 \pm 1/2\lambda)$ with
$\epsilon_l = E_C l e^{-l/\lambda}$.
Clearly, addition of the second, third, etc. particle
to the system requires the same energy $\mu_{l,+}(0)$.
For this reason, the ground
state corresponds to a regular superlattice of ECP in the
array; the period of the superlattice changes abruptly from
$l$ to $l-1$ at $\bar{\mu} = \mu_{l,+}(0) = \mu_{l-1,-}(0)$,
see Fig.~\ref{phd}a.
This simple picture of the ground state
is valid if one takes into account
the interaction of neighboring ECP
only (see Eq.(\ref{ham})).
The interaction of next-nearest neighbors
will lead to new ground states \cite{Hubbard}
in exponentially narrow regions
$|\bar{\mu} -  \mu_{l,+}(0)| \sim \epsilon_l e^{-l/\lambda}$
near the points $\mu_{l,+}(0)$.

For small but finite Josephson coupling, $E_J \ll U_0(l)$,
each bond of the length $l \pm 1$ can be considered as
{\it a mobile excitation} above the ground state
where all the bonds have the length $l$.
Since a shift of an ECP
to the neighboring island of the array leads to a shift
of the excitation by $\mp l$ lattice cells, the excitations have
{\it fractional charge} $\mp 2e/l$ \cite{fraccharge}.
Tunneling of excitations decreases
the energy by an amount $E_J$ per excitation.
Since addition  (subtraction) of one Cooper pair to (from)
the array is accompanied by the creation of
$l$ excitations, the threshold chemical potentials are given by
$\mu_{l,\pm}(E_J) = \mu_{l,\pm}(0) \mp l E_J$.

In the superlattice regime under consideration (see Fig.~\ref{phdqual})
the fluctuations of displacement of ECP are small and we can again expand the
interaction term of the Hamiltonian
(\ref{ham}) around the average distance $l$ between the particles.
We introduce the (integer) deviation $n_j = x_j - x_{j-1} - l$ of the
distance between neighboring ECP from $l$,
and the canonically conjugated operator
$\varphi_{j}$, so that $p_j = \varphi_{j+1} -\varphi_{j}$.
The number of particles $N$ can be expressed via $n_j$,
$N = (L - \sum n_j)/l$.
In terms of the new variables the Hamiltonian (\ref{ham})
can be written as
\begin{equation}
H = - E_J \sum_{j} \cos(\varphi_{j+1} - \varphi_j - a)
   +  U_0(l) \sum_{j} (n_j + \delta\mu)^2,
\label{hamharmint}
\end{equation}
where $\delta\mu =  \lambda (\bar{\mu} - \epsilon_l) /\epsilon_l$.
This Hamiltonian
formally coincides with the Hamiltonian of
1-D Josephson array with {\it on-site} Coulomb interactions,
which has been extensively studied in Refs.
\cite{esi,Freer,esilong,Bradley}.

The boundary of the commensurable phase can be determined
from the comparison of the ground state energies of the Hamiltonian
(\ref{hamharmint}) in two subspaces of states with
$\sum n_j = 0$ and with $\sum n_j = \pm 1$
(the sign coincides with the sign of
$\epsilon_l - \bar{\mu}$ in (\ref{hamharmint})).
In the limit $E_J \to 0$ the ground states in these subspaces
are given by $\Psi_0 = | 0,0,...\rangle$  and
$\Psi_{\pm 1} = N^{-1/2} \sum_j |0,...,n_j=\pm 1,0,...\rangle$.
Evaluating the energies of these states up to the third order in $E_J$
(see Ref.~\cite{Freer}) we obtain
\begin{equation}
  \mu_{l,\pm}(E_J) = \epsilon_l \pm
 \left\{ \frac{\epsilon_l}{2 \lambda}
- l E_J \left[ 1- \frac{E_J}{8U_0 (l)} -  \frac{E_J^2}{32U_0^2 (l)} \right]
\right\},
\label{phbsmallEj}
\end{equation}
for $E_J \ll U_0 (l)$, see Fig.~\ref{phd}b.
Note that the term  linear in $E_J$ coincides with that obtained above
from a simple argument.

With increasing $E_J$ the range of the chemical potential
corresponding to commensurable phase decreases and both phase boundaries
tend to the critical point, $\mu_{l,\pm}(E_J^{(cr)}) \rightarrow \epsilon_l$,
see Fig.~\ref{phd}b.
Clearly, the true behavior near the critical point
cannot be described by perturbation theory of finite order.
To extent the perturbative approach, an extrapolation
to infinite order in $E_J$ was proposed~\cite{Freer}.
Unfortunately, this (somewhat speculative)
extrapolation fails to converge to a critical point
for the 1-D system.
To determine the behaviour near the critical point $E_J^{(cr)}$
one can map the Hamiltonian (\ref{hamharmint}) (with
$\bar{ \mu } = \epsilon_l$) onto 2-D XY model~\cite{Bradley}.
The parameter $(2 U_0(l)/E_J)^{1/2}$
plays a role of dimensionless temperature $k_B T/J$ in the XY model.
The point of the Kosterlitz-Thouless transition \cite{KT} corresponds to
$E_J^{(cr)} \cong 2.5 U_0 (l)$, see Fig.~\ref{phd}b.
Below the transition temperature ($E_J > E_J^{(cr)}$)
spin-spin correlations in the XY model decay algebraically with distance.
The Josephson array shows superconducting properties:
the Josephson current is inversely proportional to $L$. \hfill
It scales as~\cite{Bradley}
 $1 + c \sqrt{E_J - E_J^{(cr)}}$
at $E_J \to E_J^{(cr)} + 0$ (here $c$ \hfill is \hfill non-
\noindent
universal constant).
Above the transition temperature
($E_J < E_J^{(cr)}$) the
correlations in the XY model decay exponentially.
Near the critical point ( $E_J \to E_J^{(cr)} - 0$)
the coherence length is given by
$\xi = \exp \{ -b [E_J^{(cr)}/(E_J^{(cr)} - E_J)]^{1/2} \} $,
where $b \simeq 2$.
As a result, the Josephson current through 1-D array decays as
$\exp( - L / l \xi)$
signalling the formation of the insulating phase.

In the insulating phase, the energy gap for the mobile excitations
scales as~\cite{esi,Freer} $\xi^{-1}$. For this reason, the
boundary of insulating phase near the critical point is given by
\begin{equation}
  \mu_{l,\pm}(E_J) = \epsilon_l \pm \alpha l U_0(l) / \xi,
\label{phbcrit}
\end{equation}
with $\alpha \sim 1$.
The upper curve in Fig.~\ref{phd}b corresponds to an extrapolation
of Eq.(\ref{phbcrit}) from the neighborhood of the
critical point to lower values of $E_J$
($1.56 U_0(l) < E_J < E_J^{(cr)}$). This extrapolation is
joined to the extrapolation of the perturbative result
(\ref{phbsmallEj}) to infinite order in $E_J$
(drawn in the range $E_J < 1.56 U_0(l)$).
A smooth connection of the two curves occurs for $\alpha \simeq 1.73$
(and $b=2$).

For completeness, we present the mean-field result for
the phase boundary~\cite{esi,esilong},
\begin{equation}
  \mu_{l,\pm}(E_J) = 	\epsilon_l \pm
			\frac{\epsilon_l}{2 \lambda}
			\sqrt{ 1- \frac{2E_J}{U_0 (l)} }.
\label{phbmeanfield}
\end{equation}
Although this expression coincides with  Eq.~\ref{phbsmallEj}
to first order in $E_J$, the overall shape of the boundary
(Fig.~\ref{phd}b)
and the critical value of the Josephson energy $E_J^{(cr)}$
differ considerably from the results discussed above.
The reason for the failure of the mean-field approach
is the absence of long-range order in one dimension.

We return now to the consideration of the superconducting phase
and concentrate on the case of small Josephson coupling, $E_J \ll U_0(l)$.
The commensurable phases with $\nu = 1/l$ and $\nu = 1/(l-1)$
are separated by a narrow superconducting region,
$\mu_{l,+}(E_J) < \mu < \mu_{l-1,-}(E_J)$,
of the width $(2l-1) E_J$.
In this region the ground state of the array can be viewed as a gas
of fractionally charged mobile excitations (bonds of length  $l-1$).
The excitations interact with each other via a contact potential
$U_{exc}(x)
= 2 U_0 (l-1) \delta_{x,0}$.
This is effectively a hard-core interaction
provided that  $E_J \ll U_0(l)$. From the consideration of the
hard-core particles (bonds of length $l-1$) on a lattice
formed by the bonds of length $l$ we obtain the chemical potential
$\bar{\mu}$ and the Josephson current $I_J$ as
functions of the concentration
$\nu = 1/(l-q)$ of ECP,
\begin{equation}
\bar{\mu} =  \mu_{l,+}(0) - E_J \{ (l-q) \cos \pi q + \pi^{-1} \sin \pi q \},
\label{muexc}
\end{equation}
\begin{equation}
I_J = \frac{4eE_J}{\hbar L} \frac{\sin \pi q}{l-q}
\frac{\Phi}{\Phi_0},
\label{IJexc}
\end{equation}
where the parameter $q$ takes on values in the range $0 < q < 1$.
These dependences are presented in Fig.~\ref{phd}c.

Finally,
we discuss effects which are specific for finite size
circular arrays.
If the size $L$ of the array is commensurable with the spacing $l$
of the superlattice of ECP, the Josephson current is exponentially small
in the insulating phase. However, if $L/l$ is not an integer,
a number of {\it residual} mobile excitations
remain in the array in the insulating phase,
$\mu_{l,-}(E_J) < \bar{\mu} < \mu_{l,+}(E_J)$.
In the lower part of this range,
$\mu_{l,-}(E_J) < \bar{\mu} < \mu_{l,0}(E_J)$,
there are $m = mod(L, l)$ residual excitations
(bonds of the length $l+1$) in the ground state.
One ECP enters into the array at $\bar{\mu} = \mu_{l,0}(E_J)$.
As a result, for  $\mu_{l,0}(E_J) < \bar{\mu} < \mu_{l,+}(E_J)$
a new ground state will contain $l-m$ residual excitations
(bonds of the length $l-1$).
The threshold chemical potential
is given by
$\mu_{l,0}(E_J) = [ m \mu_{l,-}(E_J) + (l-m) \mu_{l,+}(E_J) ] / l$.
Since each excitation contribute an amount
$I_{1CP} = (4 \pi e E_J/ \hbar L^2) (\Phi/\Phi_0)$
to the Josephson current
(cf. Eq.~(\ref{Ijharm}) with $\nu = 1/L$),
the latter shows a jump at $\bar{\mu} = \mu_{l,0}(E_J)$.

There is clearly a need for future investigations,
such as an analysis of effects of disorder
due to the offset charges (potential disorder)
and due to non-uniformity of the Josephson coupling
(kinetic disorder).

I would like to thank Yu.V. Nazarov and
A. Shelankov for a set of useful discussions
and P. Hadley for critical reading of the manuscript.
The financial support of the European Community
through HCM Fellowship ERB-CHBI-CT94-1474
is gratefully acknowledged. This work is also a part
of INTAS-93-790 project.


\begin{references}
\bibitem{remark1} For simplicity we omit logarithmic factor in this estimate,
see Eq.(\ref{nubose}).
\bibitem{remfermions} Formally, $H$ can be maped also on the Hamiltonian
of Fermions on the lattice, see Ref.~\cite{Lieb}.
\end{references}
\end{document}